\documentclass[aps,prl,twocolumn,showpacs,amsmath,amssymb,superscriptaddress,floatfix]{revtex4}
\usepackage{graphicx,epsf}
\usepackage{bm}

\begin{document}

\title{Adittional levels between Landau bands due to vacancies in graphene: 
\\ towards a defect engineering}

\author{A. L. C. Pereira}
\author{P. A. Schulz}
\address{Instituto de F\'\i sica, Universidade Estadual de Campinas - UNICAMP,  C.P. 6165, 13083-970, Campinas,
Brazil}

\date{\today}

\begin{abstract}
We describe the effects of vacancies on the electronic properties of a graphene sheet in the presence of a perpendicular magnetic field: from a single defect to an organized vacancy lattice. An isolated vacancy is the minimal possible inner edge, showing an antidotlike behavior, which results in an extra level between consecutive Landau levels. Two close vacancies may couple to each other, forming a vacancy molecule tuned by the magnetic field.  We show that a vacancy lattice introduce an extra band in between Landau levels with localization properties that could lead to extra Hall resistance plateaus.
\end{abstract}



\maketitle


\section{I. Introduction}

Graphene is one of the topical stages for research in condensed matter in the last few years. Since the seminal work describing the obtention of single graphene layers \cite{geim1}, a great amount of surprising results was obtained \cite{geim_rev,rev_mod_phys}. 
Of special interest for the present work is the observation of the quantum Hall effect in graphene \cite{novoselov,zhang1} and the related discussion in the  origin of the disorder effects and the role of vacancies in this context.
Vacancies are natural defects in the graphene lattice and can also be externally induced by ion-beam irradiation. Among the theoretical studies dealing with the effects of vacancies on the electronic structure of graphene, the main focus until now is on situations in the absence of magnetic field \cite{castroneto1,hjort,castroneto2008,brey2}. Within the experimental framework, images identifying atomic vacancies in graphene have been reported \cite{iijima}. However, although the subjacent motivation driven by the observation of the quantum Hall effect in graphene even at room temperature \cite{novoselov2}, the results on the graphene electronic structure modifications due to vacancies in the presence of magnetic field are much sparser \cite{castroneto2}.

In this paper we consider the effects of vacancies in graphene lattices when a perpendicular magnetic field is applied, showing that vacancies give origin to extra (localized) states between Landau levels (LLs). 
When a sufficiently high density of vacancies is considered, we verify that these states can couple to each other, producing additional bands with non-negligible density of states (DOS) between LLs. 
This problem has not been exhaustively explored in conventional semiconductor-based two-dimensional electron gases either. The line shape of Landau bands, as well as the experimentally observed inter-LLs DOS \cite{tu}, is a problem posed since the very beginning in the history of the quantum Hall effect, but many issues still remain  without a conclusive answer \cite{rolf,sarma}.
The first available experimental observations of the LLs DOS in graphene \cite{eva1,eva2} seem indeed to show a non-vanishing background DOS between LLs and, for the measurements on the surface of highly oriented pyrolytic graphite, unidentified DOS structures between LLs were reported \cite{eva1}. 

The coupling between the wave functions of two vacancies is verified to be determined by the distance between them and by the magnetic length tuned by the magnetic field. When considering densities of vacancies, we introduce the vacancies in a periodic pattern, which is equally spaced in the graphene lattice. Strikingly, from the investigation of the localization properties of the additional bands formed between the LLs due to these vacancy-patterns, we envisage that they could lead to interesting features in magneto-transport measurements, such as extra Hall resistance plateaus. A random distribution pattern is expected for vacancies seen as natural disorder sources or formed after irradiation with ion-beams. However, due to the advantageous electronic properties of graphene, such as room-temperature ballistic transport \cite{geim_rev,rev_mod_phys}, microstructuring efforts have experienced a fast recent development after initial concept devices. Two groups have recently reported Coulomb blockade fingerprints in graphene quantum dots with typical dimensions in the tens of nanometers range \cite{geim3,ensslin}. Furthermore, the first attempts of tailoring graphene sheets with scanning probe microscope lithography have also been reported in the literature \cite{biro,maan}. Therefore, envisaging new structures is a widespreading goal for graphene based applied investigations. Rethinking previously tested structures and geometries on a different research context has always been a motivation for several groups. A recent example is the concept of an antidot lattice on graphene \cite{graphene_antidot}. Taking advantage of these characteristics, together with the fact that microstructured graphene may lead to a working device without capping layers, an even further step could be proposed: vacancies may be thought not as defects, but may be introduced on purpose by following a previous design. From this point of view, vacancies are the minimal possible inner edges or, in other words, antidot units.
	
\section{II. Vacancies in the Lattice Model}

The results shown here are obtained within a tight-binding model with nearest neighbors hoppings in a honeycomb lattice, a framework that adequately grasps the electronic properties of graphene monolayers. We calculate the energy spectrum of graphene layers with the mentioned vacancies in the presence of perpendicular magnetic fields.  In this model, we use the following Hamiltonian:

\begin{equation}
H = \sum_{i} \varepsilon_{i} c_{i}^{\dagger} c_{i}
+ t  \sum_{<i,j>} (e^{i\phi_{ij}} c_{i}^{\dagger} c_{j} + e^{-i\phi_{ij}}
c_{j}^{\dagger} c_{i})
\end{equation}

\hspace{-\parindent}where $c_{i}$ is the fermionic operator on site $i$.
The magnetic field $B$ is included by means of a Peierls' substitution in hopping parameter ($t$$\approx$2.7eV for graphene): $\phi_{ij}= 2\pi(e/h) \int_{j}^{i} \mathbf{A} \! \cdot \! d \mathbf{l} \;$. In the Landau gauge, $\phi_{ij}\!=\!0$ along the $x$ direction and $\phi_{ij}\!=\pm \pi (x/a) \Phi / \Phi_{0}$
along the $\mp y$ direction, with $\Phi / \Phi_{0}=Ba^{2}\sqrt{3}e/(2h)$  ($a$=2.46{\AA} is the lattice constant for graphene). On-site disorder is introduced for the sake of LL bands broadening in the DOS calculations \cite{ana_2008}: white-noise fluctuations are included by sorting uncorrelated orbital energies within $\varepsilon_{i} \leq |W/2|$.


\begin{figure}[b]
\vspace{0.1cm}
\centerline{\includegraphics[width=7.6cm]{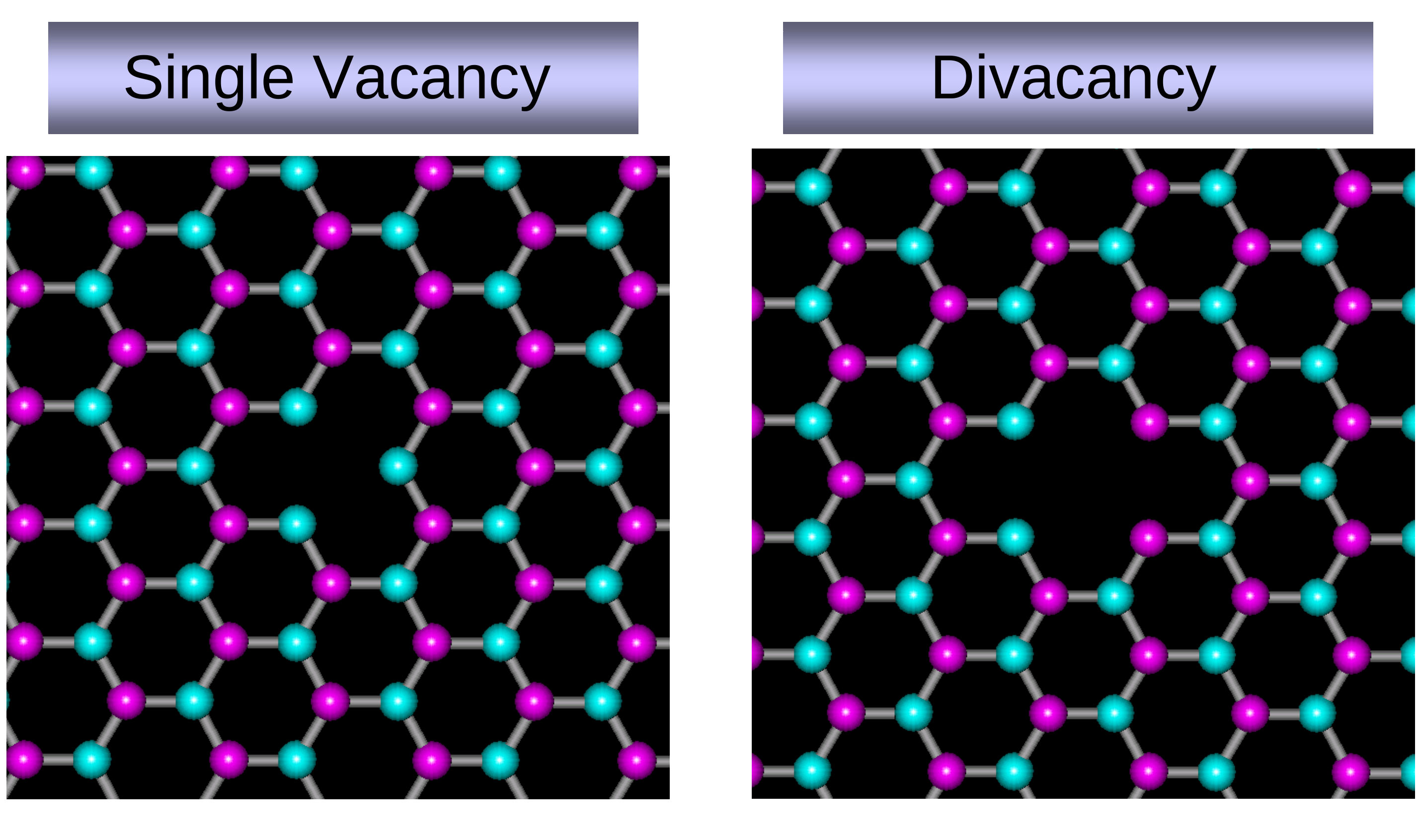}}
\vspace{-0.2cm}
\caption{(Color online) Representation of two typical vacancies in the graphene lattice: Single vacancy (also called mono-vacancy) and divacancy (corresponding to two neighbor atomic vacancies).} 
\end{figure}


We start by dealing with graphene systems containing isolated monovacancies or divacancies, as depicted in Fig. 1. Different colors for the carbon atoms represented in Fig. 1 indicate the two different sublattices. Vacancies are simulated by setting the hopping parameters to zero and the on-site energy at the defect site equals to a large value outside the energy range of the DOS. The atomic relaxation contributions to the atomic geometry of the mono and divacancies in graphene are not considered. 
Reconstruction and rebounding around vacancies do not modify qualitatively the results we showed in this work.  The main ingredient here is that vacancies are deep defects showing states that are strongly localized around them, as expected from numerical simulations \cite{castroneto2008,lee}. Furthermore, there are experimental evidences that the atomic positions of the carbon atoms in graphene do not reconstruct around a mono-vacancy \cite{iijima}.
In our model, the vacancies are introduced in graphene rectangular lattices of up to 8.000 atomic sites, which are repeated by means of periodic boundary conditions.

\section{III. Vacancy states between Landau levels}


\begin{figure}[b]
\vspace{-0.5cm}
\includegraphics[width=8.3cm]{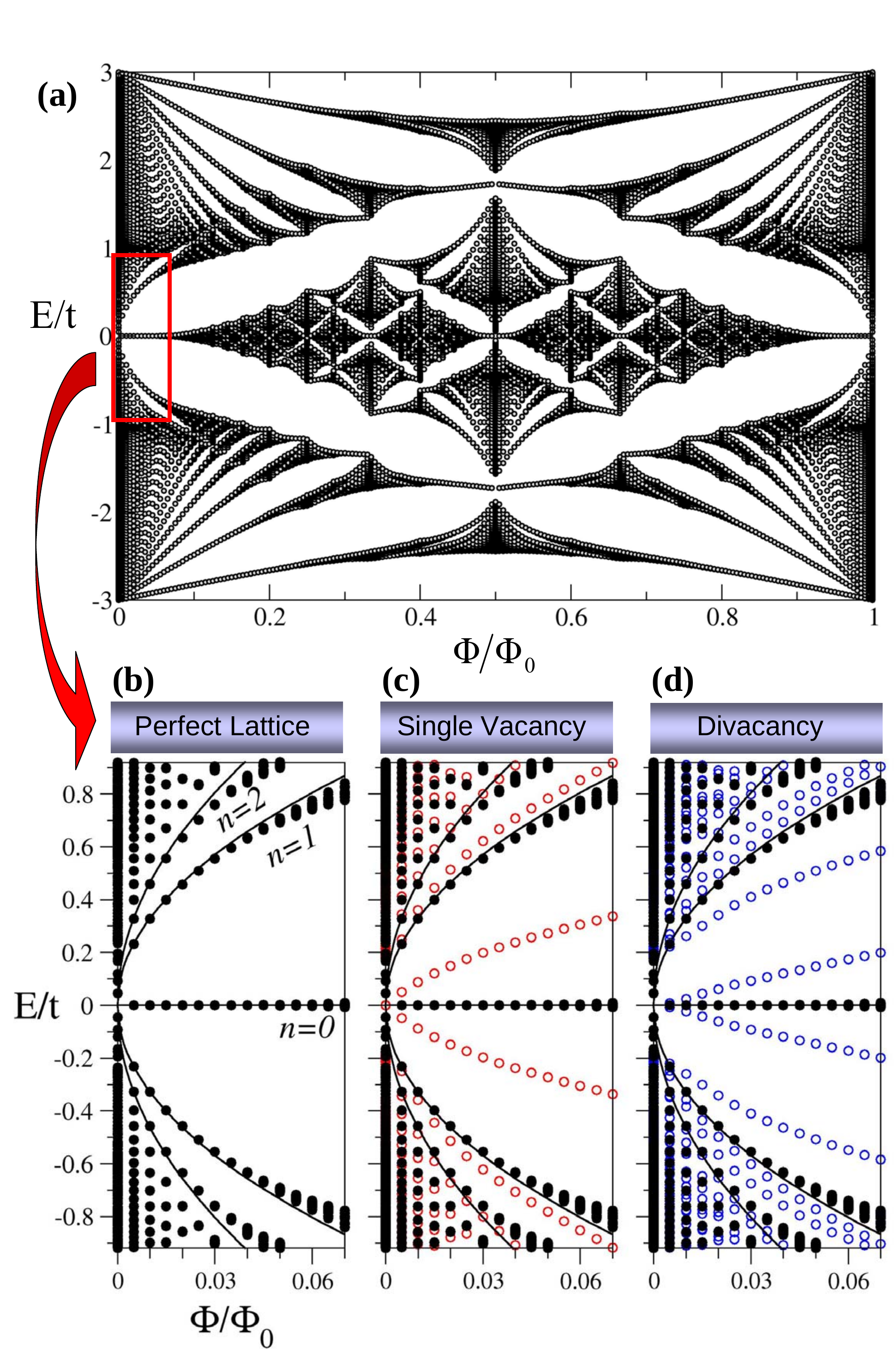}
\caption{ (Color online) {\bf (a)} Energy spectrum as a function of magnetic flux for the hexagonal lattice. Rectangular region identifies the continuum limit of the spectrum, corresponding to the LLs of a monolayer graphene. {\bf Bottom:} Energy-magnetic flux spectra for {\bf (b)} a perfect lattice, {\bf (c)} lattice with a single-vacancy and {\bf (d)} lattice with a divacancy. Continuum lines show the energy dependencies with $\sqrt{\Phi/\Phi_0}$ (i.e., with $\sqrt{B}$) of the $n$=0,$\pm$1,$\pm$2 graphene LLs, for guidance. Vacancies in the lattice introduce states with energies between the LLs, which are shown as open circles, for both vacancy types considered.}
\end{figure}


The complete electronic spectrum as a function of magnetic flux (the well-known Hofstadter-type spectrum) obtained for the perfect honeycomb lattice (no disorder) is depicted in Fig. 2(a). For the purpose of the present work, we focus on the low-flux limit, which is  identified by the rectangular area and showed in more detail in Fig. 2(b). In this limit, with energy window around the Dirac point, the LLs are well defined. Superposed to the spectra of Figs. 2(b), (c) and (d), continuum lines indicate (for guidance) the energy dependencies with $\sqrt{nB}$ of the graphene LLs with indices $n$=0,$\pm$1,$\pm$2. For higher magnetic fluxes, the lattice effects are already seen as a broadening of the highly degenerate LLs and a deviation of $n\neq 0$ LLs from the continuum lines. 

In Fig. 2(c) one can observe the evolution of the states introduced by the absence of an atom (single vacancy) in the graphene sheet: a defect state appears in each energy range between consecutive LLs (open circles). Lowering the magnetic field brings the defect states to the center of the spectrum (zero mode energy) as already discussed in comprehensive studies in the absence of magnetic fields \cite{castroneto2008}. We observe that the energies of the states introduced by the vacancy are increased with magnetic flux, however, these two states do not follow an energy dependency with $\sqrt{B}$ (followed by the LLs) nor have a linear increase with $B$. This result can be interpreted as the evolution of an edge state. Indeed, a vacancy is a minimal inner zigzag edge. The energies of external edge states are lowered with magnetic field, while states from inner edges should follow the opposite fate \cite{sivan}.
A divacancy is observed to introduce two states in the energy ranges between consecutive LLs [open circles in Fig. 2(d)]. These two states, as for the previous case, follow a different energy dependence from the LLs. From the point of view of an edge picture, a divacancy mixes zigzag and armchair characters. The inclusion of relaxations around the vacancies is not expected to modify qualitatively the physical picture shown here.

\section{IV. Two Close Vacancies: Coupling of the Localized States}

If two single vacancies are introduced far from each other in the lattice, two degenerated states are introduced in each inter-LL energy window at the same energy positions shown in Fig. 2(c) for only one vacancy. However, a different scenario shows up if the two vacancies are introduced close to each other as revealed in the energy-magnetic-flux spectra shown in Figs. 3(b-f). Five inter-vacancies distances are considered: $D$$=$5$a_{cc}$, $D$$=$6$a_{cc}$, $D$$=$7$a_{cc}$ [schematically represented in Fig. 3(a)], $D$$=$8$a_{cc}$, and $D$$=$9$a_{cc}$, where $a_{cc}$=$1.42$\AA $\,$ is the carbon-carbon distance, i.e., the atomic spacing between two neighbor carbons in the graphene lattice. 


\begin{figure}[b]
\includegraphics[width=8.7cm]{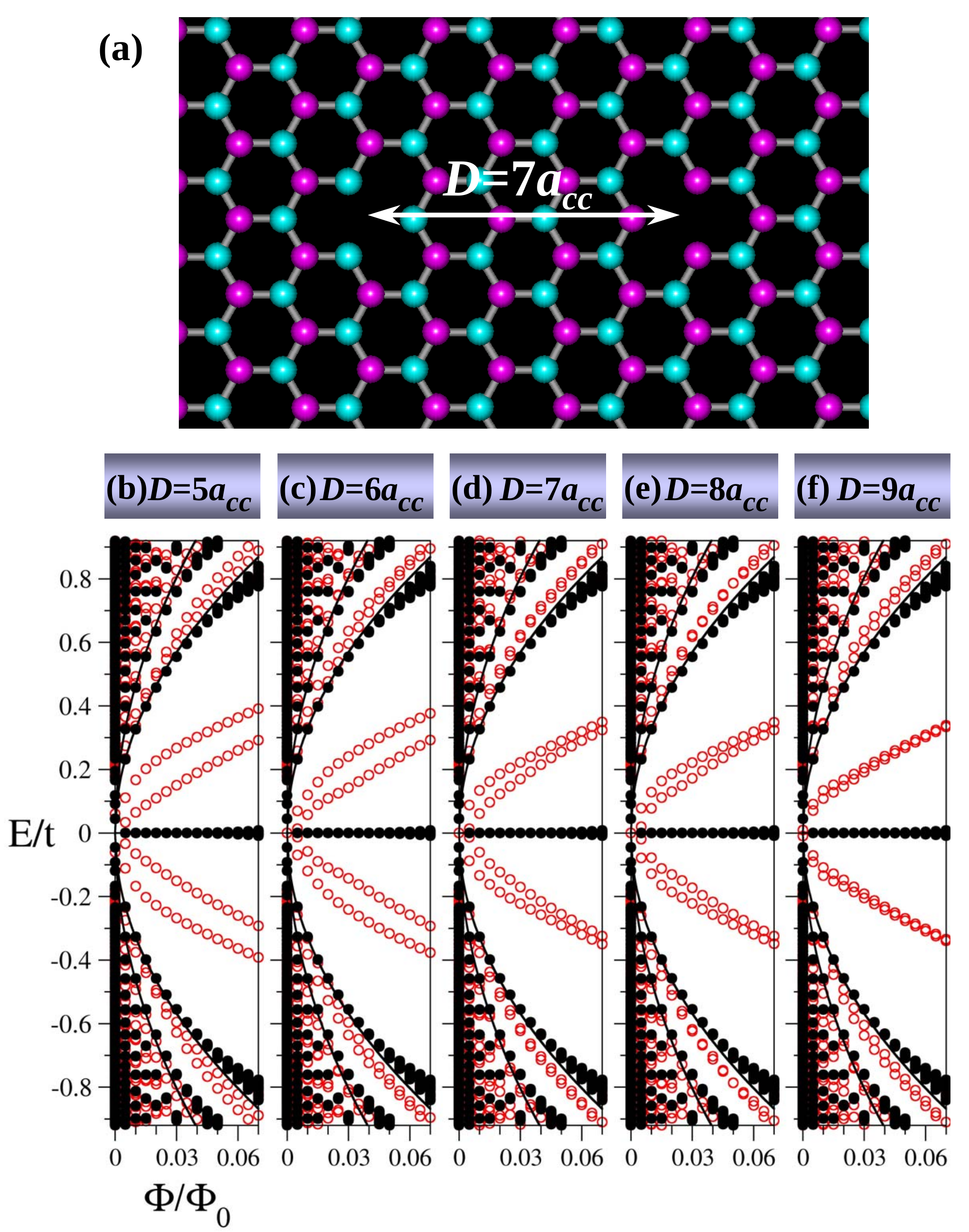}
\vspace{-0.6cm}
\caption{ (Color online) {\bf (a)} Representation of two close vacancies in the graphene hexagonal lattice, with a separation between them of $D$ = 7$a_{cc}$, where $a_{cc}=1.42$\AA $\,$ is the distance between two neighbor carbon atoms. {\bf Bottom:} Energy-magnetic-flux spectra for lattices with two close vacancies, separated by increased distances: {\bf (b)} 5$a_{cc}$, {\bf (c)} 6$a_{cc}$, {\bf (d)} 7$a_{cc}$, {\bf (e)} 8$a_{cc}$, and  {\bf (f)} 9$a_{cc}$. Two states due to the two vacancies appear between LLs (open circles) and the energy splitting between these states increases for decreasing distance $D$ or decreasing magnetic fields.}
\label{Fig3} \end{figure}


Under the main features identified in the spectra of Fig. 3 are the very energy splitting of the defect states and the evolution of this splitting with magnetic field. One expects that these levels split if there is a hybridization between the states localized around the vacancies. For an inter-vacancy distance of $D=9$$a_{cc}$ [Fig. 3(f)], the splitting observed is becoming small and the two vacancy-related states almost collapse in energy at high magnetic fields. Hence, one can see in Fig.3 that the energy splitting decreases with increasing vacancy distances or increasing magnetic field. 
These main features observed are robust independently of whether the two vacancies are situated in the same sublattice or in different sublattices. In fact, for the inter-vacancies distances $D$$=$5$a_{cc}$, $D$$=$7$a_{cc}$, and $D$$=$8$a_{cc}$ shown here, the two vacancies are in different sublattices, while for the cases $D$$=$6$a_{cc}$ and $D$$=$9$a_{cc}$, the two vacancies are located in the same sublattice.
The zero mode states (in the absence of magnetic field) associated to single vacancies are localized around the absent atoms \cite{castroneto2008}. In the presence of a magnetic field, this localization could be seen from the point of view of a different scale,  given by the magnetic length $l_B$=$\sqrt{\hbar/eB} \approx 257 \,$\AA$/\sqrt{B(Tesla)}$; increasing the magnetic field will then lead to a spatial shrinking of the wave function. This shrinking leads to a quenching of the hybridization of the vacancy states, which will start to be effective only for $l_B \lesssim D/2$. 

\vspace{-0.62cm}

\section{V. Vacancy Superlattices}

\subsection{A. Extra bands in the DOS}

We then consider higher vacancy concentrations, introduced in a periodic pattern. As we discuss in Sec. I, having in mind a promising and bearable scenario of microscopic manipulation, a vacancy lattice - the ordered counterpart of a disordered network - is worth to be considered.   
In Fig. 4(a) the DOS of a graphene lattice without vacancies is compared with the DOS of a lattice containing a rectangular arrangement of 12 vacancies, corresponding to a vacancy concentration of  $x$=$0.21\%$. The dimensions of the lattices considered here in Fig. 4 are 96$\times$60=5760 atomic sites, arranged in 96 zig-zag chains, each of them containing 60 sites. 
Here a slight white-noise on-site disorder is also introduced ($W/t=0.1$) in order to introduce a broadening in the highly degenerate LLs, as well as to permit an unambiguous calculation of the localization properties.

The focus of the present work is on the extra bands created between the LLs [clearly shown in Fig. 4(a)]. For the magnetic flux considered, $\Phi/\Phi_0=2/96\approx 0.02$, the bands due this density of vacancies are already comparable to the bulk LLs. It should be kept in mind that, for a fixed concentration of vacancies, the relative weight of these extra bands further increases with diminishing magnetic fields, since the number of states in each extra band is equal to the number of vacancies in the lattice, while the LL degeneracy is proportional to the magnetic field. In fact, the LL degeneracy is given by $N_{total} \times \Phi/\Phi_0$, the total number of atomic sites considered times the magnetic flux.    
Another point to be noticed is that, for inter-vacancies distances corresponding to this vacancy concentration of $x$=$0.21\%$, the associated energy levels are still degenerated (when the white-noise disorder is not included). This can be verified by inspecting that the widths of the LLs and of these extra-bands are basically the same, induced by the white-noise disorder.
The higher concentration showed here, of $0.42\%$ [in Fig 4(b)], imposes a vacancy lattice with distances between vacancies of $D \approx 18 a_{cc}$ in both lattice directions. 
Therefore, vacancy (antidot) lattice effects are expected in the electronic spectrum for magnetic fluxes at which the ``bulk" graphene LLs are still well defined. This indeed is the origin of some non-trivial modulation observed for the DOS between the higher LLs (not shown here).


\begin{figure}[t]
\begin{center}
\includegraphics[width=8.3cm]{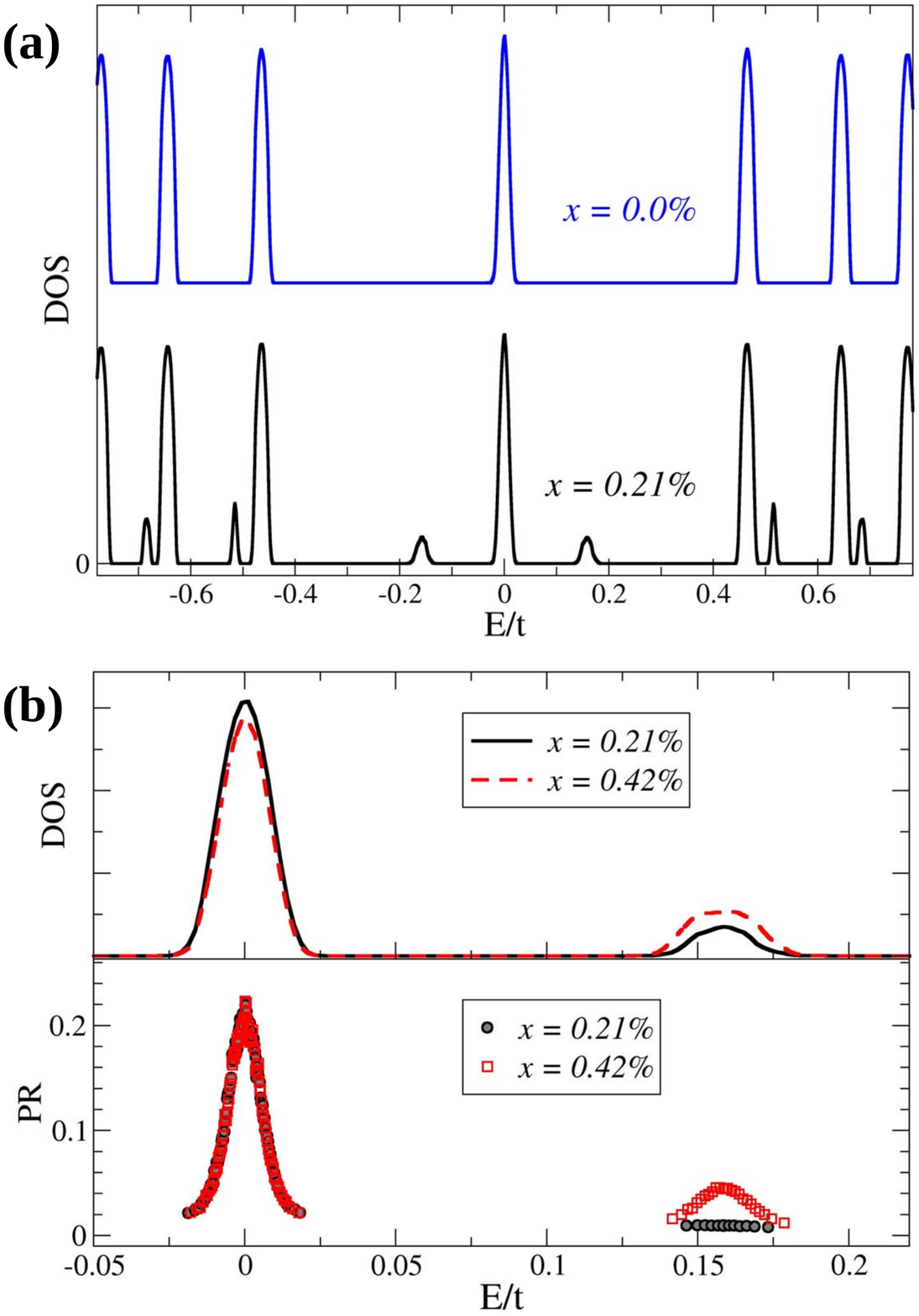}
\vspace{-0.3cm}
\end{center}
\caption{ (Color online) {\bf (a)} Density of states for a graphene lattice without vacancies compared with a lattice containing a vacancy concentration of $x$=$0.21\%$ shifted vertically for clarity. The formation of extra bands between LLs, due to the vacancies, is evident. LLs are broadened by a white-noise disorder with $W/t$=0.1 and $\Phi/\Phi_0=2/96\approx 0.02$. {\bf (b)} DOS and participation ratio (PR) for states within the zero-energy LL and the first vacancy-related band. Increasing vacancy concentration ($x$) leads to a modulation of the PR within the extra band similar to that of the LL. }
\label{Fig4} \end{figure}


\vspace{-0.1cm}

\subsection{B.  Delocalization within vacancy bands}

We inspect the localization properties of the states within the extra vacancy bands by means of the participation ratio (PR) \cite{thouless},

\begin{equation}
PR = \frac{1}{N \sum_{i=1}^{N}|\psi_{i}|^{4}},
\end{equation}

\hspace{-\parindent}where $\psi_{i}$ is the amplitude of the normalized wave function on site $i$, and $N$ is the total number of lattice sites.
The PR gives the proportion of $N$ sites over which
the wave function is spread. In this way, the PR for a localized state will tend to zero in the thermodynamic limit.

In Fig. 4(b), attention is focused only on the Landau bands $n$=$0$ and $n$=$+1$. We see from the PR calculation that if the concentration of vacancies is low ($x=0.21\%$), all the states in the vacancy bands are localized. On the other hand, if the intervacancy distances diminish, at certain magnetic-flux range (roughly when $l_B > D/2$) these vacancy states are effectively coupled and eventually become delocalized. Strikingly, the PR for a vacancy concentration of $x=0.42\%$ shows a strong modulation for the vacancy related band, similar to the ones for the Landau bands \cite{ana_2008}. This behavior signalizes that a vacancy lattice could be identified in magnetotransport measurement, possibly by extra quantum Hall plateaus.

\section{VI. Discussion and Conclusions}

For both vacancy concentrations shown in Fig. 4, all the vacancies were introduced in the same sublattice. However, we emphasize that we also tested configurations (not shown here) where the vacancies are equally distributed between the two sublattices (zero degree of uncompensation), and exactly the same effect seen in Fig. 4 is observed, i.e., a vacancy-band is formed in the DOS and, after a critical concentration, the PR is modulated, indicating a delocalization of the states in the middle of the vacancy-band. In the absence of magnetic fields, it was shown \cite{castroneto2008} that for lattices containing vacancies with an increasing degree of uncompensation between sublattices, an increasing energy gap appears in the DOS around zero energy, with localized states in the middle of the gap. However, the magnetic field introduces a  highly degenerate Landau level exactly at zero energy, which substantially modifies the scenario. Having in mind that for the concentrations and magnetic fluxes considered no changes are noticed due to sublattice uncompensation, we have that here, in the same way that for the case of two vacancies, the effect we are describing does not show dependence on the degree of uncompensation between sublattices. Nevertheless, the uncompensation effects on the limit of B$\to$0, i.e, on the transition from well-defined LLs to B=0, deserve further investigations.

We would like to mention that comparisons with experimental conditions are not hindered by the drastic limitations imposed by the periodic boundary conditions, which are further constrained by the commensurability related to the magnetic flux. For the largest lattice used in the results shown here (with 8000 sites being 400 zig-zag chains, each with 20 sites in the results from Figs. 2 and 3), the lowest possible magnetic flux is still $\Phi / \Phi_{0}=2/400=0.005$, corresponding to $B\approx400$T, a value that is one order of magnitude above attainable experimental setups \cite{kim}.
Intervacancy coupling, the main underlying cause of the present findings, is significant whenever $l_B \geq D/2$. This condition implies, for $B=40T$, an intervacancy distance $D \leq 80$\AA, which corresponds to a vacancy concentration of the order of $x$=0.1$\%$. An intervacancy distance of $80$\AA  is also much lower than the estimations of both: mean-free path and coherence length in graphene samples.

It should be emphasized that our results are not qualitatively dependent on the precise position in the energy of the vacancy-related levels. As stated in Sec. II, the basic ingredients for building ``vacancy molecules" or ``vacancy superlattices" are wave functions localized around the absent atoms that can be tuned by the magnetic field. The only condition in these additional levels is that they lie far apart in energy from the LLs, a very robust condition. Besides that, an important fingerprint is the qualitatively different evolution of the states with magnetic field, deviating from the usual $\sqrt{B}$ dependence for lateral LLs.

In summary, our main findings are listed below.
(i) Vacancies in graphene introduce states between LLs, which tend to higher energies with increasing magnetic fields, consistent with the behavior of inner edge states. These states could be assessed by the usual switching of the magnetic field or gate voltages in quantum Hall measurements, but are not expected to reveal any fingerprint in these measurements because of their strong localized character.
(ii) However, sufficiently near vacancies will lead to the coupling of these localized states, leading to  ``vacancy molecules". The bonding between these vacancies can be tuned by the magnetic field. 
(iii) A periodic vacancy lattice introduces well-defined extra bands between the LLs with a non-negligible DOS. They do not affect the transport properties only at low concentrations. However, in the low magnetic field limit or higher concentrations, when vacancy states can be effectively coupled, vacancy superlattice delocalized states will cross the system and directly associated magneto-transport features can be expected.

ALCP acknowledges the support from FAPESP. PAS received partial support from CNPq.


\end{document}